\documentclass[12pt]{iopart}

\def\Journal#1#2#3#4{{#4} {\it #1} {\bf #2}, #3 }
\def\ac{\overline{\alpha}}
\def\bc{\overline{\beta}}
\def\gc{\overline{\gamma}}
\def\dc{\overline{\delta}}
\def\ec{\overline{\epsilon}}
\def\rc{\overline{\rho}}
\def\mc{\overline{\mu}}
\def\tc{\overline{\tau}}
\def\pc{\overline{\pi}}
\def\kc{\overline{\kappa}}
\def\nc{\overline{\nu}}
\def\s5{\sqrt{5}}

\begin{document}

\title{An exhaustive classification of aligned Petrov type D purely magnetic perfect fluids}

\author{Norbert Van den Bergh\footnote{e-mail:
norbert.vandenbergh@ugent.be} and Lode Wylleman\footnote{Research assistant supported by
the Fund for Scientific Research Flanders(F.W.O.), e-mail:
lwyllema@cage.ugent.be}}

\address{Faculty of Applied Sciences TW16, Gent University, Galglaan 2, 9000 Gent, Belgium}

\begin{abstract}
We prove that aligned Petrov type D purely magnetic perfect fluids
are necessarily locally rotationally symmetric and hence are all
explicitly known.
\end{abstract}

\pacs{0420}



\section{Introduction}
There has been a recent surge of interest~[1--10] in purely
gravito-magnetic space-times, which are defined as (non-conformally
flat) space-times for which a time-like congruence $v$ exists such that the gravito-electric part of the
Weyl-tensor with respect to $v$ vanishes:
\begin{equation}\label{E_ab}
E_{ac}\equiv C_{abcd} v^b v^d=0 .
\end{equation}
The resulting space-times are then necessarily~\cite{McIntosh} of
Petrov type I or D, with the congruence $v$ being uniquely defined
for Petrov type I and with $v$ an arbitrary timelike vector in the
plane of repeated principal null directions for Petrov type
D~\cite{Barnes, Lozanovski5}. Whereas large and physically important
classes of examples exist for purely gravito-electric space-times
(for example all the static space-times 
are purely electric), little
information is available for the purely magnetic ones. This is
particularly true for the vacuum solutions (with or without
$\Lambda$ term), where no purely gravito-magnetic solutions are
known at all. This has lead to the conjecture that purely
gravito-magnetic vacua do not exist~\cite{McIntosh}, but
so far this has only been proved when the Petrov type is
D~\cite{McIntosh}, or when the timelike congruence $v$ is
shearfree~\cite{Haddow}, non-rotating~\cite{Trumper,VdB1},
geodesic~\cite{VdB2}, or satisfies certain technical generalisations of these
conditions~\cite{Ferrando0, Carminati2}. 
In \cite{Ferrando1, Ferrando2} the non-existence of
irrotational purely magnetic models was generalised to space-times
with a vanishing Cotton tensor.\\

For perfect fluid models on the other hand, with
\begin{equation}
G_{ab}\equiv R_{ab}-\frac{1}{2}R g_{ab}= (w+p) u_a u_b +p g_{ab},
\end{equation}
a proof has recently~\cite{Wylleman} been given that, when the pressure is constant and the vorticity is zero
(the so called anti-Newtonian universes), no space-times exist for which (\ref{E_ab}) holds with respect to $u$.
For non-constant pressure little work so far has been done on the algebraically general case,
with the exception of
\cite{Wylleman2}, in which an example is given of a non-rotating, non-accelerating magnetic perfect fluid of Petrov type I,
and research so far has been concentrated on the `aligned' Petrov
type D solutions (see however also \cite{Bonnor}). These are defined
as perfect fluid models for which the fluid velocity $u$ is aligned
with the repeated principal null directions $k$ and $l$ of the Weyl
tensor. For a purely magnetic space-time it is then clear that the
property (\ref{E_ab}) holds with respect to $u$, see
also~\cite{Lozanovski5}. Alternatively, if (\ref{E_ab}) holds with
respect to the fluid velocity, then imposing the Petrov type D
condition automatically implies that the fluid is aligned.
Remarkably all known aligned Petrov type D purely magnetic perfect
fluids are locally rotationally symmetric (LRS): this holds e.g.~in
the non-rotating case for the $p=\frac{1}{5}w$ Collins-Stewart
space-time~\cite{Collins} and the Lozanovski-Aarons
metric~\cite{Lozanovski1}, and in the rotating case, for the
stationary and rigidly rotating model of \cite{Fodor}, and for all
their LRS generalisations~\cite{Lozanovski2,Lozanovski3}. Actually
it was proved in \cite{Lozanovski2,Lozanovski3}, making use thereby
of an earlier result on the
 shear-free solutions~\cite{Lozanovski5}, that non-rotating or shear-free purely
 magnetic aligned Petrov type D perfect fluids are necessarily LRS of
 Ellis' class III or I and that the
 resulting metric forms, which could all be explicitly determined,
 thereby exhaust the LRS purely magnetic perfect fluid solutions. \\

In the present paper we go one step further, by demonstrating that
the LRS family completely exhausts the aligned Petrov type D purely
magnetic perfect fluids. We will make use of the Newman-Penrose
formalism and follow the notation and sign conventions 
of \cite{Kramer}, whereby the
Newman-Penrose equations (7.21a -- 7.21r) and Bianchi identities
(7.32a -- 7.32k) will be indicated as (np1 -- np18) and (b1 -- b11)
respectively. All calculations were carried out using the Maple symbolic algebra package\footnote{A module for manipulation of the Newman-Penrose equations can be obtained from the authors.}.

\section{Main equations}

We use a canonical type D tetrad, with
\begin{equation}\label{e:psis} \Psi_0=\Psi_1=\Psi_3=\Psi_4=0 ,\end{equation} with
 the condition (\ref{E_ab}) being expressed as
\begin{equation}\label{e:psi2}
\overline{\Psi_2}=-\Psi_2
\end{equation}
Choosing a boost in the $(k,\ell)$
plane such that the fluid velocity $u=(k+\ell)/ \sqrt{2}$ and introducing $S=w+p$ as a new variable, one has
\begin{equation}\label{e:phis}
\Phi_{00}=\Phi_{22}=2 \Phi_{11}=\frac{S}{4}\ \textrm{and}\ R=4 w-3
S.
\end{equation}
Substitution of the conditions
(\ref{e:psis},\ref{e:psi2},\ref{e:phis}) in the Bianchi identities
$(b1), (b2), \Im(b3), (b4),$ $(b3)-(b6), \Im(b6), (b7)+\overline{(b2)}, (b8)-(b5), (b9), (b10),
(b11)$ one obtains straightforwardly the algebraic restrictions
\begin{eqnarray} \lambda = \sigma = \nu+\overline{\kappa}+3
\overline{\tau}+ 3 \pi = 0, \label{e:lsnu} \\
18\Psi_2 (\rho-\rc+\mc-\mu)-S
(\rho+\rc-\mu-\mc+2\epsilon+2\ec-2\gamma-2\gc)=0 \ \label{bi3min6}
\end{eqnarray}
together with
\begin{eqnarray}
\delta \Psi_{{2}}  =   -\frac{1}{12}\left( 2 \overline{\alpha } +2
\beta-\kappa -\overline{ \nu } +\overline{ \pi}  +\tau
 \right) S+ \left( \kappa+3\tau \right) \Psi_{{2}} \\
 D \Psi_2  =   \frac{3}{2}(\rho+\rc)
 \Psi_2+\frac{S}{8}(\rho-\rc+\mu-\mc) \\
 \Delta \Psi_2  =   -\frac{3}{2}(\mu+\mc)
 \Psi_2-\frac{S}{8}(\rho-\rc+\mu-\mc) \\
\delta S   =   -12 \kappa\,\Psi_{{2}}+\left(2 \overline{ \alpha} +2
\beta +\kappa -
\overline{\pi}  \right) S  \\
\delta w  =  -12 \kappa\,\Psi_{{2}}+ 2 \left(\overline{ \alpha} + \beta-\overline{\pi} - \tau  \right) S \\
D w  = \frac{1}{2}(D-\Delta)S+\frac{S}{2}(\rho+\rc-\mu-\mc+2 \gamma+2 \gc) \\
\Delta w  = -\frac{1}{2}(D-\Delta)S+\frac{S}{2}(\rho+\rc-\mu-\mc-2
\epsilon -2\ec) \end{eqnarray}
as well as an expression for $(D-\Delta) S$,
\begin{equation}\label{e:bi3}
(D-\Delta) S   =  -18(\rho-\rc) \Psi_2+\frac{1}{2}(\rho+\rc-\mu-\mc-2\gamma-2\gc)S ,
\end{equation}
which in combination with (\ref{bi3min6}) could be used for further
simplification of $Dw$ and $\Delta w$: this however turns out to be
disadvantageous when applying e.g.~the $[\delta, \, D+\Delta]$
commutator to $w$.

The expressions for shear, vorticity, expansion and acceleration, simplified by means of (\ref{e:lsnu}), are presented in the appendix.\\

The key observation is now that the combination $3 (np7) +(np10)+\overline{(np2)}-3 \overline{(np16)}$ factorises as follows:
\begin{equation}\label{e:main}
(\alpha+\bc)(2\kc+3 \pi+3 \tc)=0.
\end{equation}
In the next paragraphs we will discuss the resulting cases
separately: in section 3 it is shown that $2\kc+3 \pi+3 \tc=0$ leads
to local rotational symmetry, while in section 4 we prove that no solutions
exist when $2\kc+3 \pi+3 \tc\not = 0$.

\section{$2\kc+3 \pi+3 \tc=0$}

In this case one obtains from (\ref{bi3min6})
\begin{equation}
\kappa = \nc = -\frac{3}{2} (\tau+\pc)
\end{equation}
after which $(b5)$ can be solved for $\beta$:
\begin{equation}
\beta = -\ac+\tau+\pc
\end{equation}

Then however $\overline{(np2)}-(np10)$ implies
\begin{equation} \tau+\pc =0,\end{equation}
which means that the vorticity vector and the spatial gradient of
$w$ are parallel with the vector $k-\ell$ . From the equations
$(np1),(np3),(np7),(np9),(np11),(np13),(np14)$ one easily obtains then
\begin{equation}\label{e:spaceeqs}
\eqalign{
\delta \rho & = \pc (\rc-\rho) \\
D \rho & = \rho (\rho+\epsilon+\ec) +\frac{S}{4} \\
\dc \mu & = \pi (\mc-\mu) \\
\Delta \mu & = -\mu(\mu+\gamma+\gc) -\frac{S}{4} \\
\dc \pi & = -\pi(\pi+2 \alpha) \\
D \pi & = \pi (\ec-\epsilon) \\
\Delta \pi & = \pi(\gc-\gamma)
}
\end{equation}
while the combination
$\overline{(np18)}-(np15)-(np5)-\overline{(np4})$ results in the
relation
\begin{equation}\label{e:useful}
\delta (\gamma+\gc -\epsilon -\ec) +\pc (\gamma+\gc-\epsilon-\ec+\mu-\rc) = 0 .
\end{equation}
Using (\ref{bi3min6})
to simplify the expression which results by acting with the
commutator $[\dc,\delta]$ on $w$, we find a further factorisation,
\begin{equation}\label{e:maincase1}
S(\rho+\rc-\mu-\mc)(\rho-\rc+\mu-\mc)=0.
\end{equation}
As we can ignore the Einstein spaces ($S=0$), for which the
non-existence of purely magnetic Petrov type D solutions was
demonstrated in \cite{McIntosh}, solutions are either vorticity-free
($\rho-\rc+\mu-\mc=0$) or are rotating and have
$\rho+\rc-\mu-\mc=0$. In the latter case $\theta_{33}$ (see
appendix) is the only
possible non-zero component of the expansion tensor and below we show that actually $\theta_{ab}=0$.\\

The first case is the one treated in \cite{Lozanovski2}. The application of the $[\delta,\, D+\Delta]$ commutator to $w$ results then in
\begin{eqnarray}
\rc-\mu = \rho-\mc \\ \delta(\rc-\mu+\gamma+\gc-\epsilon-\ec) =0,
\end{eqnarray}
with which (\ref{e:useful}) simplifies to
\begin{equation}
\delta(\rc-\mu)+\pc (\rc - \mu +\ec+\epsilon-\gc-\gamma) = 0
\end{equation}
and hence, using (\ref{e:spaceeqs}),
\begin{equation}
\pc (\mu-\rc+\gamma+\gc -\epsilon -\ec)=0.
\end{equation}
If $\pi\not = 0$ then (\ref{bi3min6}) results in
$\Psi_2(\mu-\mc)=0$, after which the real part of $(np12)$ would
imply $\Psi_2=0$. Therefore $\pi, \tau, \kappa, \nu, \lambda,
\sigma$ and hence also $\delta R$, are all 0: the solutions are then
LRS according to the theorem by Goode and
Wainwright~\cite{GoodeWainwright}. The  resulting metrics are
explicitly described
in \cite{Lozanovski2}.\\

When the vorticity is non-zero, necessarily
\begin{equation}\label{e:case1bdef}
\rho+\rc-\mu-\mc=0,
\end{equation}
and (\ref{bi3min6}) simplifies to
\begin{equation}\label{refbimin6X}
18 \Psi_2 (\rho-\mu)+S(\gamma+\gc-\epsilon-\ec)=0 .
\end{equation}
Applying next the $[\delta, \, \Delta]$ and $[\delta, \, D]$
commutators to $w$, making use of (\ref{e:bi3}) and eliminating
$\delta(\gamma+\gc-\epsilon-\ec)$ from the resulting equations by means of
(\ref{e:useful}), results in the following linear system for $\delta
\mu$ and $\delta \rc$:
\begin{eqnarray}
\fl S \delta(\mu) +(36 \Psi_2-S) \delta \rc +108 \pc(\rho-\rc)\Psi_2
+\pc (4\gamma+4\gc-4\epsilon-4\ec+\rho-\mc-5\rc+5\mu)S  = 0 \nonumber \\
\ \\
\fl S \delta \mu -(S+12 \Psi_2) \delta \rc -36 \pc (\rho-\rc)
\Psi_2+\pc (\mu-\mc+\rho-\rc)S  = 0.
\end{eqnarray}
Solving this system for $\delta \mu$ and $\delta \rc$ allows one to
apply the $\delta$ operator to the defining equation
(\ref{e:case1bdef}), from which one obtains
\begin{equation}\label{e:case1bb}
\pc (18(\rho-\mu)\Psi_2 +S(\rc-\mu))=0.
\end{equation}
Again observe that $\pi\not = 0$ is not allowed: equations
(\ref{e:case1bdef}, \ref{refbimin6X}, \ref{e:case1bb}) imply then $\Psi_2(\rho-\mu)=0$
and hence also $\rc-\mu=0$,
in contradiction with the assumptions that the vorticity is non-zero, namely $\rho-\rc \not =0$,
that $\Psi_2$ is imaginary and $S$ is real.\\

We conclude that $\pi=0$: just as for the non-rotating case the conditions
of the Goode-Wainwright theorem are then satisfied and solutions are
LRS.  They are therefore shearfree~\cite{Lozanovski3} and the metric
forms are discussed in detail in \cite{Lozanovski3, Lozanovski5}.

\section{$2\kc+3 \pi+3 \tc\not =0$}

From (\ref{e:main}) one now obtains $\beta = -\ac$, which implies
the restrictions $\sigma_{13}+\omega_2=\sigma_{23}-\omega_1=0$ on
the shear and vorticity (see appendix). From the expressions of the
latter it is clear that $\omega_1=\omega_2=0$ is not allowed: this
would imply $\tau+\pc=0$ and hence, using ($b5$), also $\kc
\Psi_2=0$, which takes us back to the previous section. We can
therefore fix the tetrad by requiring $\omega_2=0$. Writing
$\omega_1=\sqrt{2}\omega$ ($\overline{\omega}=\omega$) this can be
expressed as
\begin{equation}
\tau =-\pc- i \omega,
\end{equation}
after which ($b5$) simplifies to
\begin{equation}
\kappa= \frac{i \omega}{6 \Psi_2}(9\Psi_2+S).
\end{equation}
From $(np16)-\overline{(np7)}$ one obtains now
\begin{equation}
\delta \omega = -i \omega^2 -2\omega  (\pc+\ac),
\end{equation}
which, when substituted in $(np2)$, yields
\begin{equation}\label{e:pi}
\pi = \frac{i \omega}{6 S \Psi_2} (S^2+3 S \Psi_2 +54 \Psi_2^2),
\end{equation}
the substitution of which in $(np7)$ or $(np16)$ implies
\begin{equation}
\omega^2 ( 5 S^2 + 81 \Psi_2^2) =0.
\end{equation}
Without loss of generality we can therefore assume
\begin{equation}
S=\frac{9 i}{\sqrt{5}} \Psi_2.
\end{equation}
Herewith (\ref{e:pi}) becomes
\begin{equation}
\pi=\frac{7+i \sqrt{5}}{2 \sqrt{5}}\omega ,
\end{equation}
while (\ref{e:bi3}) and (\ref{bi3min6}) reduce to a pair of algebraic equations, which allow one to express the real parts
of $\epsilon$ and $\gamma$ as functions of $\rho,\rc,\mu$ and $\mc$:
\begin{eqnarray}
\Re\left( (8 \s5 + 9 i) \rho  + (4 \s5 -31 i) \mu  + 8 \s5 \epsilon \right) =0  \label{bi3bis} \\
\Re\left( (4\s5-31 i)\rho+(8 \s5+9 i) \mu +8 \s5 \gamma) \right)=0. \label{bi3min6bis}
\end{eqnarray}
We now can solve $(np3)$ and $(np9)$ for $D \omega$, $\Delta \omega$: expressing that the latter derivatives are real,
results ---with the aid of (\ref{bi3bis}) and (\ref{bi3min6bis})--- in a homogeneous system for $\rho, \mu$ and their conjugates,
\begin{eqnarray}
\Re\left( 3(9\s5 -139 i) \rho +(97 \s5 + 133 i) \mu \right) = 0, \label{e:eq7} \\
\Re\left( (97 \s5 + 133 i) \rho +3 (9 \s5 -139 i) \mu \right) =0  , \label{e:eq8}
\end{eqnarray}
from which one obtains
\begin{equation}\label{e:muu}
\mu = \frac{94 i \s5}{367} \rho -\frac{604+499 i \s5}{1101}\rc .
\end{equation}
A tedious calculation also allows one to solve $(np11, np13,np5+\overline{np4},np15-\overline{np18})$
for $\delta \rho, \delta \rc, \delta \mu, \delta \mc$. Simplifying these results with (\ref{e:muu}) yields
\begin{eqnarray}
\delta \rho = -\frac{1044 \s5-1135 i }{1835}\omega\rho+\frac{2129\s5-50 i}{1835}\omega \rc \\
\dc \rho = 3 \frac{4609 \s5-285 i}{3670} \omega \rho -\frac{3256 \s5 + 3475 i}{1835} \omega \rc \\
\delta \mu = \frac{10412 \s5 + 10145 i}{5505} \omega \rho -\frac{9753 \s5+22525 i}{3670}\omega \rc \\
\dc \mu = -\frac{1436\s5 -3515 i}{1835}\omega \rho+\frac{3223 \s5 -2950 i}{5505} \omega \rc .
\end{eqnarray}
Using the latter expressions to evaluate the $\delta$ derivative of (\ref{e:muu}) eventually yields
\begin{equation}
(2670895 i-5018217\s5)\rho-3(1352555 i-5051814\s5)\rc = 0,
\end{equation}
from which one obtains $\rho=0$ and hence also $\mu=0$. Herewith $(np12)$ implies $\Psi_2=0$, in contradiction with the assumption that
the Petrov type is D.

\section{Conclusion}
When, for an aligned Petrov type D purely gravito-magnetic perfect fluid, a canonical null-tetrad is chosen such that equations (\ref{e:psis},\ref{e:psi2},\ref{e:phis}) hold, we proved that necessarily
$\kappa=\lambda=\sigma=\nu=\tau=\pi=\alpha+\bc=\delta R = 0$. Solutions are then locally rotationally symmetric and belong to one of the classes discussed in detail in \cite{Lozanovski2, Lozanovski3}: they either have $\rho-\rc=\mc-\mu$ and are non-rotating with non-vanishing shear, or they have $\rho+\rc =\mu+\mc$ and are rotating and shearfree.

\section{Appendix}
Choosing an orthonormal tetrad such that $\delta\equiv(e_1 - i
e_2)/\sqrt{2}$, $D \equiv(e_3+e_4)/\sqrt{2}$ and $\Delta \equiv
(e_4-e_3)/\sqrt{2}$ ($e_4=u$ being the fluid velocity) and taking into account the simplifications (\ref{e:lsnu}), the components of the fluid kinematical quantities are given by the
following expressions:

(expansion tensor)
\begin{eqnarray}
\theta_{12} = 0 \\
\theta_{13}+i \theta_{23} = (\alpha+\bc+2\pi+2\tc)/\sqrt{2} \\
\theta_{11} = \theta_{22} = (\mu+\mc-\rho-\rc)/(2\sqrt{2})\\
\theta_{33}= (\epsilon+\ec-\gamma-\gc)/\sqrt{2}
\end{eqnarray}

(acceleration vector)
\begin{eqnarray}
\dot{u}_1+ i \dot{u}_2 = -\sqrt{2}(\pi+\kc+2\tc)\\
\dot{u}_3 = (\epsilon+\ec+\gamma+\gc)/\sqrt{2}
\end{eqnarray}

(vorticity vector)
\begin{eqnarray}
\omega_1+i \omega_2 = \frac{i}{2} (\alpha+\bc - 2 \tc -2 \pi)/\sqrt{2}\\
\omega_3 = \frac{i}{2}(\rho-\rc+\mu-\mc)/\sqrt{2} .
\end{eqnarray}

\end{document}